\documentstyle[12pt]{article}
\def\be{\begin{equation}}
\def\ee{\end{equation}}
\def\ba{\begin{array}{c}}
\def\ea{\end{array}}

\begin{document}

\titlepage

\begin{center}


{\Large \bf Solvable simulation of a double-well problem

in ${\cal PT}-$symmetric quantum mechanics
 }

\vspace{1.truecm}



{\bf Miloslav Znojil}

 Department of Theoretical Physics,

 Institute of Nuclear Physics,
Academy of Sciences,

 250 68
\v{R}e\v{z}, Czech Republic

e-mail: znojil@ujf.cas.cz

\end{center}



\vspace{0.5truecm}

\section*{Abstract}

Within the framework of ${\cal PT}-$symmetric quantum mechanics
of bound states (which works with parity-pseudo-Hermitian
Hamiltonians $H = {\cal P} H^\dagger {\cal P}$ and real spectra)
we mimic some effects of the double-well structure of potentials
by a pair of $\delta$ functions with mutually complex conjugate
strengths. The model is solvable by the standard matching
technique and exhibits several interesting features. We observe
an amazingly robust absence of any ${\cal PT}-$symmetry
breaking. A quasi-degeneracy which occurs in the high-energy
domain is interpreted as a manifestation of certain ``quantum
beats".

 \vspace{0.5truecm}

PACS 03.65.Fd   03.65.Ca  03.65.Ge 03.65.Bz

\newpage

\section{Introduction}

In current textbooks, many phenomena (like, e.g., vibrational
spectra) and methods (like, e.g., perturbation expansions) of quantum
mechanics are best illustrated by the one-dimensional Schr\"{o}dinger
equation for bound states in a real and symmetric well $V(x) =
V^\ast(x) = V(-x)$,
\begin{equation}
\left[ -\,\frac{d^{2}}{dx^{2}} + V(x) \right ] \,\psi (x)=E\psi (x)
\,.
 \label{schrod}
\end{equation}
In such a setting, the transition of the so called ${\cal
PT}-$symmetric quantum mechanics (say, in its form proposed by
Bender and Boettcher \cite{BB}) may most easily be illustrated
by an inclusion of an asymmetric and purely imaginary additional
potential in the same equation,
 \be
 V(x) = V_S(x) + i\,V_A(x)\,, \ \ \ \ V_S(x) = V_S^\ast(x)=V_S(-x)
 \,, \ \ \ \ V_A(x) = V_A^\ast(x)=-V_A(-x)\,.
 \label{PTS}
 \ee
The exactly solvable examples abound \cite{ES,ES2}. As their most
elementary and transparent sample we may recollect the harmonic
oscillator in $D$ dimensions~\cite{ptho} with
 \be
 V(x) = \frac{\ell(\ell+1)}{(x-i\,c)^2} + B\,(x-i\,c)^2\,,
\ \ \ \ c > 0\,.
 \label{HOP}
 \ee
Here, the centrifugal barrier is regularized and proportional to
$\ell = (D-3)/2+m$ in the $m-th$ partial wave.

The studies of models (\ref{PTS}) also often involve much more
complicated potentials, which are not exactly solvable.  An
encouragement is provided by their more immediate applicability
as well as by the efficiency of many semiclassical \cite{BBjmp},
perturbative \cite{mytri} and/or purely numerical techniques
\cite{Handy}. In the pioneering paper, Caliceti et al
\cite{Caliceti} paid attention to the very unusual imaginary
cubic ${\cal PT}-$symmetric anharmonicity.  Within sophisticated
perturbation theory, she (together with her co-authors)
succeeded in showing that in spite of the non-Hermiticity of the
Hamiltonian, the spectrum of energies may remain real.

Buslaev and Grecchi \cite{BG} returned to the more standard, quartic
${\cal PT}-$symmetric anharmonic oscillator
 \be
 V(x) = \frac{\ell(\ell+1)}{(x-i\,c)^2} + B\,(x-i\,c)^2 + D\,(x-i\,c)^4
 \,,
\ \ \ \ c > 0\,\,
 \label{BGP}
 \ee
with a similar motivation stemming from field theory~\cite{Seznec}.
They were the first who demonstrated that the bound-state energies of
a non-Hermitian, ${\cal PT}-$symmetric model (\ref{BGP}) may coincide
with the spectrum of a certain {\em Hermitian} double-well problem.
Their sophisticated and explicit, Fourier-type equivalence
transformation between the Hermitian and non-Hermitian partners may
be now better understood in the language
of~A.~Mostafazadeh~\cite{Mostafazadeh}.

Let us note that the presence of the centrifugal term in eq.
(\ref{BGP}) played an important technical role in ref. \cite{BG},
while a similar term has been absent in all the original studies of
the imaginary cubic forces \cite{BBjmp,Caliceti}. Curiously enough,
only a re-introduction of this term in the generalized model
 \be
 V(x) = \frac{\ell(\ell+1)}{(x-i\,c)^2} + B\,(x-i\,c)^2 +
 i\,C\,(x-i\,c)^3\,,
\ \ \ \ c > 0\,
 \label{CGM}
 \ee
seems to have opened the way towards the rigorous proof of the
reality of the spectrum of the imaginary cubic oscillators
\cite{DDT}. In the light of this proof, the special regular case of
eq. (\ref{CGM}) with the vanishing $\ell(\ell+1)=0$ is not
exceptional at all. {\em On the contrary}, the inclusion of a strong
repulsion $\ell(\ell+1)\gg 1$ in eq. (\ref{CGM}) seems instructive
and productive in having paved the way towards the recent
clarification of the applicability of the current $1/\ell$ technique
in the ${\cal PT}-$symmetric context in ref.~\cite{somarem}. Thus, we
shall assume that $\ell$ is large in all the potentials of the type
(\ref{HOP}), (\ref{BGP}) or (\ref{CGM}). At this point, we get quite
close to the subject of our present short communication since the
implementation of the $1/\ell$ technique proved {\em entirely
different} in the Hermitian and ${\cal PT}-$symmetric models, and we
intend to re-analyze the latter case via certain simplified models.

\section{Double wells}

All three  ${\cal PT}-$symmetric potentials (\ref{HOP}) --
(\ref{CGM}) offer a very good testing ground for the comparison of
the Hermitian and ${\cal PT}-$symmetric calculations. Firstly, they
may serve as an elementary illustration of the $1/\ell$ technique in
the Hermitian case because their Hermitian versions emerge simply in
the limit $c \to 0$. Secondly, one immediately notices that such a
step is not {\em mathematically} trivial since their centrifugal
barriers become strongly singular. Fortunately, this does not play
any role at all because simultaneously, we have to ``shrink" the full
real axis of the coordinates to the mere positive ``radial" half-axis
\cite{ptho}. A fully rigorous discussion of this point is available
in ref. \cite{BG} and enables us to re-interpret our differential
Schr\"{o}dinger equation (\ref{schrod}) with the {\em real and
constrained} coordinates $x=x_{Hermitian}=x_{(c = 0)}>0$ as the
standard central bound-state problem on half-axis. In such a context,
the perturbative $1/\ell$ recipe (as reviewed, e.g., in paper
\cite{jednadel} and/or in many other references which are quoted
therein) is virtually trivial. One simply imagines that the potential
$V(x)$ possesses an absolute minimum at some real and positive
coordinate $R>0$ \{ for example, in eq. (\ref{HOP}) we get the {\em
unique} minimum at $R = [{\ell(\ell+1)/B}]^{1/4}$, etc \}. In the
vicinity of this minimum, the shape of the potential $V(x)$ may be
approximated by an appropriate harmonic oscillator well, $V(R+\xi)
\approx const.+ \omega^2\xi^2 + {\cal O}(1/R)$.  At this point one
discovers that the quality of the latter ``zero-order" approximation
{\em increases quickly} with the growth of $\ell$. Moreover, the
expansion of the binding energies and wave functions in the powers of
$1/R$ remains feasible and exhibits very good convergence properties
in practice~\cite{jednadel}.

After a return to the non-Hermitian eq. (\ref{schrod}) with $x \in
(-\infty,\infty)$, an application of the same idea requires a much
more careful analysis.  As long as all our sample potentials $V(x)$
remain smooth, analytic and confining for {\em all} the shift
parameters $c \neq 0$, our wave functions $\psi(x)$ stay analytic and
{\em normalizable} all over the lower half-plane of the complex plane
of $x$ \cite{BB}.  At the same time, the manifest non-Hermiticity of
our sample Hamiltonians at $c > 0$ leads to the necessity of a
re-interpretation of their eigenstates. In the spirit of our remarks
\cite{pseudo,mytrid} (cf. also the later reviews
\cite{Mostafazadeh,Bogdan}), one must replace the concepts of
Hermiticity by pseudo-Hermiticity while the unitarity becomes
weakened to pseudo-unitarity. As a consequence, the probabilistic
interpretation of ${\cal PT}-$symmetric quantum mechanics acquires
the forms which find their immediate guidance in the relativistic
quantum mechanics, with one of the most popular examples provided by
the Feshbach-Villars pseudo-Hermitian re-formulation of the
Klein-Gordon equation \cite{FV}.

The net consequence of all these introductory remarks is that within
the framework of our above exemplification of the ${\cal
PT}-$symmetric quantum mechanics we are {\em free} to work with {\em
any complex} coordinate $R$ of the minimum of $V(x)$. As long as we
assumed that $\ell(\ell+1) \gg 1$ is large, we discover that {there
exist {\em many} minima} $R$ generating, in principle, {\em many}
alternative large$-\ell$ expansions. For example, eq. (\ref{HOP})
with $c > 0$ gives the minimum of $V(x)$ whenever $x^4=R^4 =
\ell(\ell+1)/B$. Even in this trivial example (where we know all the
final solutions in advance!), we have no clear criterion for the
choice not only between the above-mentioned real $R =R_{(+)}=
\varrho$ (with the large $\varrho = |[{\ell(\ell+1)/B}]^{1/4}| \gg
1$) and its negative partner $R_{(-)}= -\varrho$ but also between the
two new purely complex extremes at $R =R_{(lower)}= -i\,\theta$ and
$R=R_{(upper)} =R^*_{(lower)}= +i\,\theta$ (with the same real
parameter $\theta =\varrho \gg 1$). Moreover, the transition to the
potentials (\ref{BGP}) and (\ref{CGM}) (both with $B=0$ for
simplicity) gives the respective rules $R^6 = \ell(\ell+1)/(2D)$ and
$i\,R^5 = 2\ell(\ell+1)/(3C)$ so that, generically, one has to deal
with several pairs of the {\em complex} minima such that
 \be
 V(R+\xi) \approx  \Omega^2\,\xi^2, \ \ \ \ \ \ \ R = R_{(\pm)}= \pm
 \varrho
 - i\,\theta, \ \ \ \ \ \ \ \ \ \   \Omega^2 =
  \Omega^2_{(\pm)}=  \omega^2  \pm i\,\eta
 \label{dvemin}
 \ee
where we only know that $|R| \gg 1$. In such a setting, we may choose
the shift of the axis in such a way that $c \equiv \theta$, {\em
without} causing any change in the spectrum of course.

The fully exhaustive discussion of the particular cubic example
(\ref{CGM}) may be found in ref.~\cite{somarem}. For the simpler
quartic oscillator (\ref{BGP}), the same large$-\ell$ construction
would lead to the vanishing and non-vanishing $\theta$ at the
positive and negative couplings $D> 0$ (occurring, e.g., in refs.
\cite{Simsek}) and $D < 0$ (chosen, e.g., in refs. \cite{BG,BBjpa}),
respectively. Thus, in contrast to the Hermitian case where the real
position of the absolute minimum of $V(x)$ is unique in the majority
of cases of practical interest, the regularization mediated by the
shifts $c \neq 0$ leads, in accordance with the pattern
(\ref{dvemin}), to the most frequent occurrence of the two symmetric
minima in $V(x)$ at once. Now we may conclude our introductory
considerations by a declaration that {\em any} ${\cal PT}-$symmetric
double well (\ref{dvemin}) with the {\em complex} (and
complex-conjugate) strengths $\Omega^2_{(\pm)}$ has not been found
solvable in our preceding paper \cite{somarem}, and that it offered
the main motivation for our forthcoming considerations, therefore.

\section{The model}

We shall try to simulate the effects of the general double attraction
in the following schematic square-well-like model
 \be
 V(x)=
 \left ( -\omega^2-i\,\eta \right )\,\delta(x+a) +
 \left ( -\omega^2+i\,\eta \right )\,\delta(x-a)\,
 \ \ \ \ \ \ \ x \in (-1,1)
 \ee
where, in the spirit of refs. \cite{ES2}, the forces are reduced to
the mere ${\cal PT}-$symmetric pair of the delta functions at a
distance measured by the variable $ a \in (0,1)$. This means that we
plan to solve eq. (\ref{schrod}) with the boundary conditions
 \be
 \psi(\pm 1)=0, \ \ \ \ \ \ \
 \frac{d}{dx} \psi(\pm a+0)-
 \frac{d}{dx} \psi(\pm a-0)=\left (-\omega^2 \pm i\,\eta \right )\,
 \psi(\pm a)\,
 \label{matching}
 \ee
by the standard matching technique as described in textbooks
\cite{Fluegge39}. Under the usual ${\cal PT}-$symmetric normalization
convention \cite{mytrid}
 \be
 \psi(x) = \psi_S(x) + i\,\psi_A(x)\,, \ \ \ \ \psi_S(x) =
 \psi_S^\ast(x)=\psi_S(-x)
 \,, \ \ \ \ \psi_A(x) = \psi_A^\ast(x)=-\psi_A(-x)\,,
 \ee
this enables us to denote $E=\kappa^2$ in the obvious ansatz for the
wave functions on the three sub-intervals of the whole domain in
question,
 \be
 \psi(x)= \left \{
 \begin{array}{ll}
 \psi_L(x) =
 (\alpha - i\,\beta)\,\sin \kappa(x+1), \ \ & x \in (-1,-a)\\
 \psi_C(x) =
 \gamma\,\cos \kappa x + i\,\delta \,\sin \kappa x, \ \ & x \in (-a,a)\\
 \psi_R(x) =
 (\alpha + i\,\beta)\,\sin \kappa(-x+1), \ \ & x \in (a,1)\\
 \ea
 \right ..
 \ee
Moreover, as long as we have
 \be
 \psi'(x)= \left \{
 \begin{array}{ll}
 \psi'_L(x) =\kappa\,
 (\alpha - i\,\beta)\,\cos \kappa(x+1), \ \ & x \in (-1,-a)\\
 \psi'_C(x) =-\kappa\,
 \gamma\,\sin \kappa x + i\,\kappa\,\delta \,\cos \kappa x,
 \ \ & x \in (-a,a)\\
 \psi'_R(x) =-\kappa\,
 (\alpha + i\,\beta)\,\cos \kappa(-x+1), \ \ & x \in (a,1)\\
 \ea
 \right .
 \ee
the appropriate insertions in eq. (\ref{matching}) lead to the
four-by-four matrix set of equations
 \be
 \left (
 \begin{array}{cccc}
 \sin \kappa (1-a)& 0& -\cos \kappa a& 0\\
 0& \sin \kappa (1-a)& 0& -\sin \kappa a\\
  -\mu(\omega)& \nu(\eta) & \sin \kappa a& 0\\
 \nu(\eta)& \mu(\omega)& 0& \cos \kappa a
  \ea
 \right )
 \left ( \ba
 \alpha\\
 \beta\\
 \gamma\\
 \delta
 \ea
 \right ) = 0\,.
 \label{match}
 \ee
Here, $\mu(\omega) = \cos \kappa (1-a) - \omega^2\,\kappa^{-1}\,\sin
\kappa (1-a) $ and $\nu(\eta) = \eta\,\kappa^{-1}\,\sin \kappa (1-a)
$ are mere abbreviations.

We may conclude that the matching conditions may be satisfied if and
only if the secular determinant $F(\kappa)$ vanishes in eq.
(\ref{match}). This condition has the following form
 \be
 F(\kappa)=\sin\,( 2\kappa) + \frac{\omega^2}{\kappa}\,
 \left [
 \cos\,( 2\kappa) -\cos (2\kappa a)
 \right ]
+
 \frac{\omega^4+ \eta^2}{\kappa^2}\,
 \sin (2\kappa a)\,
 \sin^2[ 2\kappa(1- a)]\,
  =0
 \,.
 \label{side}
 \ee
Its main merit is its compact form, not quite expected in the light
of the previous complicated matching formula (\ref{match}). By its
structure it resembles the textbook solution of the simple square
well so that, in this sense, it is a mere implicit representation of
the spectrum itself. In fact, one might even believe that an explicit
construction of this spectrum might be possible (say, in the form of
an infinite power series in some of the parameters) but due to the
utterly elementary character of eq. (\ref{side}), one might still
deduce the majority of the relevant features of the spectrum from
this implicit definition. In order to demonstrate this expectation in
detail, the roots of eq. (\ref{side}) will be studied here via the
graphs of the function $F(\kappa)$. In this sense, we shall determine
and display the various qualitative features spectrum of the energies
$E_n=\kappa^2_n$ in a series of pictures, via their presentation in
the form of the sequences of the nodal zeros $\kappa_n$ of the
function~$F(\kappa)$.

\section{Discussion}

The results of our study are sampled in Figures 1 - 7 at the
different sets of the couplings. In all of these cases we observed,
first of all, that the influence of the changes of $\eta$ is not too
relevant so that we have fixed, in all our pictures, $\eta = 20$ for
the sake of definiteness.

In Figure 1 we simulated the system of the two weakly attractive
wells which lie far from each other. We see that the choice of $a =
0.95$ and $\omega = 3/2$ still leads to the mere very weak
perturbation of the very well known spectrum of the  $\omega = 0$
energies in square well. The function $F(\kappa)$ oscillates very
regularly and its overall amplitude, although not quite constant,
exhibits just a small slow variation and remains almost independent
of our choice of the interval of $\kappa$ (the Figure takes $\kappa =
\sqrt{E} \in (0,15)$).

In Figure 2 we see that for the same distance of the wells, the
growth of their attractive strength up to $\omega = 15 000$ leads to
the significant growth of the average magnitude of the amplitude of
$F(\kappa)$ {\em and} to the emergence of an ``envelope" (i.e., an
auxiliary curve $F_{av}(\kappa)$ which connects the maxima or minima)
with much slower oscillations. In addition we discover an overall
asymptotic decrease of $F_{av}(\kappa)$ and certain less apparent
periodic structures when we try to connect some of the ``subdominant"
local maxima/minima of $F(\kappa)$.

Figures 3 and 4 show the effect of the move of the two wells towards
each other. Even a not too significant shortening of their distance
(to $a = 0.85$ and $a = 0.65$, respectively) shows that the period of
the envelopes becomes shorter and, in the latter case, one can see a
clear competition between several parallel envelopes which start to
overlap each other.

In order to reveal the latter pattern clearly, we omitted the low
lying levels. This enabled us to see that the interplay between
different periods remains fairly regular. Step by step, it introduces
a much more complicated pattern to the energy spectrum, anyhow. The
separate levels behave in a more and more chaotic manner, especially
when we return to the low-lying part of the spectrum in a way
illustrated in Figure 5 where we kept $a=0.65$ and weakened $\omega =
150$.

The remarkable feature of all this pattern is that our numerical
experiments never revealed a merger of the two levels followed,
presumably, by their disappearance in the complex plane. In the other
models, such a phenomenon may exist and is usually interpreted as the
so called spontaneous ${\cal PT}-$symmetry breaking \cite{Geza}.
Here, Figure 6 with the fairly small distance $a = 0.35$ (and with
the original very strong $\omega = 15 \cdot 10^3$) illustrates the
situation where one gets very close to such a possibility and where
several pairs of the real energies appear to be almost degenerate.

In our last Figure 7 where $\omega = 150$, we finally demonstrate
that the ``chaotic" character of the eigenvalues in our schematic and
exactly solvable ${\cal PT}-$symmetric point-interaction double wells
may be weakened and partially removed by a return to their weaker
strengths. We see in the picture that the characteristic pairwise
``irregular" quasi-degeneracies persist. This type of an irregularity
has a slightly different source in an ultimate slowdown of the
oscillations of the  envelopes, the ``wobbles" of which become
comparable to the nodal distances.

We may summarize that the double-well-like structure of our complex,
non-Hermitian example complies with some of our intuitive
expectations. Thus, the ``robust" reality of the spectrum or an
emergence of the quasi-degeneracy have been re-confirmed. At the same
time, a few other observations (say, of the multiple and/or
super-imposed ``beats" in the curve $F(\kappa)$) wait for a
deeper understanding and/or a generic explanation in the future.

\section*{Acknowledgments}

Work partially supported by GA AS (Czech Republic), grant Nr. A 104
8302.

\section*{Figure captions}

\noindent Figure 1. The sinus-like left-hand side curve $F(\kappa)$
 of eq.
 (\ref{side}) in the weakly perturbed square-well regime,
 i.e., at a ``large" distance  $a = 0.95$ of the ``shallow" wells
 with a weak strength $\omega = 1.5$ and with
 a (virtually irrelevant) asymmetry measure $\eta = 20$.

 \noindent Figure 2.
 Graphical determination
 of the high-lying energies $E = \kappa^2$ (at nodes of $F(\kappa)$)
 for a very strong attraction at $\omega = 15 \cdots 10^3$
 (the other parameters are the same as in Figure~1).

\noindent Figure 3. A shortening of the ``beats" of Figure 2
 after the shortening of the distance between wells to $a = 0.85$.

\noindent Figure 4. The emergence of the overlaps between multiple
``beats" at a still shorter distance $a=0.65$.

\noindent Figure 5.  A return to the lowest energy levels in Figure 4
and to the weaker attraction with $\omega = 150$.

\noindent Figure 6. The quasi-degeneracy of levels
 for the ``deep" wells with  $\omega = 15 \cdot 10^3$ at a comparatively
 small distance $a=0.35$.

\noindent Figure 7. Same as Figure 6, with a ``wobbling-type"
irregularity in the energy nodes for ``shallow" wells with $\omega =
150$.


\newpage


\begin{thebibliography}{00}


\bibitem{BB}  Bender C M and Boettcher S 1998 Phys. Rev. Lett. {24} 5243

\bibitem{ES}
Cannata F, Junker G and Trost J 1998 Phys. Lett. {\ A 246 } 219;

Znojil M 1999 Phys. Lett. A. 264 108;

Andrianov A A, Cannata F, Dedonder J P and Ioffe M V 1999 Int. J.
Mod. Phys. A 14  2675;

Bagchi B and Roychoudhury R 2000 J. Phys. A: Math. Gen. 33 L1;

Znojil M 2000 J. Phys. A: Math. Gen. 33 L61 and 4561;

Bagchi B, Cannata F and and Quesne C 2000 Phys. Lett. A 269 79;

L\'{e}vai G and Znojil M 2000 J. Phys. A: Math. Gen. 33 7165;

Ahmed Z 2001 Phys. Lett. A 286 231;

Znojil M and Tater M 2001 J. Phys. A: Math. Gen. 34 1793;

Tkachuk V M and Fityo T V 2001 J. Phys. A: Math. Gen. 34 8673;

Klishevich S M and Plyushchay M S 2001 Nucl. Phys. B 616 403;

Bagchi B, Mallik S and Quesne C 2002 Int. J. Mod. Phys. A 17 51;

L\'evai G, Cannata F and Ventura A 2002 J. Phys. A: Math. Gen. 35
5041;

Bagchi B and Quesne C 2002 Phys. Lett. A 300 18;

Sinha A and Roychoudhury R 2002 Phys. Lett. A 301 163;

Jia C-S, Sun Y and Li Y 2002 Phys. Lett. A 305 231

\bibitem{ES2}
Znojil M 2001 Phys. Lett. A. 285 7;

Albeverio S, Fei S M and Kurasov P 2002 Lett. Math. Phys. 59 227;

Nagasawa T, Sakamoto M and Takenaga K 2002 SUSY in QM with point
interactions, LANL arXiv hep-th/0212192

\bibitem{ptho}
Znojil M 1999 Phys. Lett. A. 259 220

\bibitem{BBjmp}
Alvarez G 1995 J. Phys.{\ A: Math. Gen. 27} 4589;

Delabaere E and Pham F 1998 Phys. Lett. A 250 25 and 29;

Bender C M, Boettcher S and Meisinger P N 1999 J. Math. Phys. 40
2201;

Delabaere E and Trinh D T 2000 J. Phys.{\ A: Math. Gen. 33} 8771;

Shin K C 2001 J. Math. Phys. 42 2513;

Nanayakkara A and Abayaratne C 2002 Phys. Lett. A 303 243

\bibitem{mytri}
Fernandez F, Guardiola R, Ros J and Znojil M 1999 J. Phys. A: Math.
Gen. 32 3105;

Bender C M and Dunne G V  1999 J. Math. Phys. 40 4616

\bibitem{Handy}
Handy C R, Khan D, Wamg Xiao-Xian and Tomczak C J 2001 J. Phys. A:
Math. Gen. 34 5593;

Bernard C and Savage V M 2001 Phys. Rev. D 64 085010;

Bender C M and Weniger E J and 2001 J. Math. Phys. 42 2167

\bibitem{Caliceti}  Caliceti E, Graffi S and Maioli M 1980 Commun. Math.
Phys. 75 51

\bibitem{BG}  Buslaev V and Grecchi V 1993 J. Phys.{\ A: Math. Gen. 26} 5541

\bibitem{Seznec}
Seznec R and Zinn-Justin J 1979 J. Math. Phys. 20 1398;

Avron J and Seiler R 1981 Phys. Rev. D 23 1316

\bibitem{Mostafazadeh}
Mostafazadeh A 2002 J. Math. Phys. 43 205, 2814 and 3944

\bibitem{DDT}  Dorey P, Dunning C and Tateo R 2001 J. Phys. A: Math. Gen. 34
5679 and L391.

\bibitem{somarem}
Znojil M, Gemperle F and Mustafa O 2002 J. Phys A: Math. Gen. 35 5781

\bibitem{jednadel}
Bjerrum-Bohr N E J 2000 J. Math. Phys. 41 2515

\bibitem{pseudo}
Znojil M 2001 ``Conservation of pseudo-norm in PT symmetric
quantum mechanics", arXiv math-ph/0104012, to appear in
Suppl. ai Rendiconti del Circ. Matem. di Palermo.

\bibitem{mytrid}
Bagchi B, Quesne C and Znojil M 2001 Mod. Phys. Lett. 16 2047

\bibitem{Bogdan}
Ram\'{\i}rez A and Mielnik B 2002 Rev. Fis. Mex., to appear (arXiv:
quant-ph/0211948)

\bibitem{FV}
Feshbach H and Villars F 1958 Rev. Mod. Phys. 30 24

\bibitem{Simsek}
Znojil M 1999 J. Phys. A: Math. Gen. 32 7419;

Bender C M, Berry M, Meisinger P N, Savage V M and Simsek M 2001  J.
Phys. A: Math. Gen. 34 L31

\bibitem{BBjpa}
Bender C M and Boettcher S 1998 J. Phys.{\ A: Math. Gen. 31} L273;

Znojil M 2000 J. Phys. A: Math. Gen. 33 4203

\bibitem{Fluegge39}
Fluegge S 1971 Practical Quantum Mechanics I (Berlin: Springer), p.
39

\bibitem{Geza}
Khare A and Mandal B P 2000 Phys. Lett. A 272 53;

L\'{e}vai G and Znojil M 2001 Mod. Phys. Lett. A 16 1973;

Ahmed Z 2001 Phys. Lett. A 282 343;

Znojil M and L\'{e}vai G 2001 Mod. Phys. Lett. A 16 2273


\end{thebibliography}
\end{document}